\documentstyle[11pt,newpasp,epsf]{article}
\newcommand{\EBV}{E$_{\rm B-V}$~}
\newcommand{\Tef}{T$_{\rm eff}$~}

\begin{document}

\title{
Model atmospheres and SEDs of \\
chemically peculiar stars.
}

\author{Yakiv V. Pavlenko}

\affil{Main Astronomical Observatory, Golosiiv woods,
Kyiv-127, 03680 Ukraine; e-mail: yp@mao.kiev.ua; 
http://www.mao.kiev.ua/staff/yp}





\setcounter{footnote}{3}


\begin{abstract}

 Procedure and results of computations of stellar model atmospheres
and spectral energy distributions are discussed.
Model atmospheres of some chemically peculiar stars are computed
taking into account detailed information about their
abundances:

--- R CrB-like stars of Teff $\sim$ 7000 K,

--- Sakurai's object (V4334 Sgr) of 4000 $<$ \Tef $<$ 7000 K

--- Przybylski's star of Teff $\sim$ 6500 K.

We show that our self-consistent approach provides a unique
possibility to investigate the temporal changes of physical
parameters of chemically peculiar stars.

Some issues of computation of model atmospheres of M and C-giants
are also considered.

\end{abstract}

\section{Introduction}

  In many aspects, the existence of the irregular 
hydrogen-deficient (Hd) variables remains puzzling so far. R CrB 
is the most known member of the post-AGB group. Sakurai's object 
(SO, V4334 Sge) provides another, extreme case of stellar 
evolution.

It has been firmly established that the most abundant elements in 
atmospheres of R CrB-like stars are helium and carbon. 
Determination of abundances in their atmospheres is possible only 
in the frame of self-consistent approach (Asplund et al. 1998). 
Still even in the case of the ``normal'' red giants that approach 
should be used. Otherwise, abundance determination results might 
be affected by significant errors ($>$ 0.2-0.3 dex, see Pavlenko 
\& Yakovina 1994 for more details).

HD~101065 (V816 Cen) presents
another case of the peculiar stellar spectrum (see Cowley et al.
2000) ---- the strongest spectral lines in the
spectrum of HD~101065 generally belong to lanthanides.

\section{Ionization-dissociation equilibrium}
In comparatively
cool (\Tef $<$ 6500 K) atmospheres of Hd stars densities of
carbon containing molecules increase (Pavlenko 2002b). Strong molecular
bands of CO, CN, C$_2$ appear in optical and IR spectra
of Sakurai's object (Pavlenko 2002b, Pavlenko \& Geballe 2002).

\section{Opacities}
\begin{figure}
\begin{center}
\epsfxsize=100mm \epsfbox{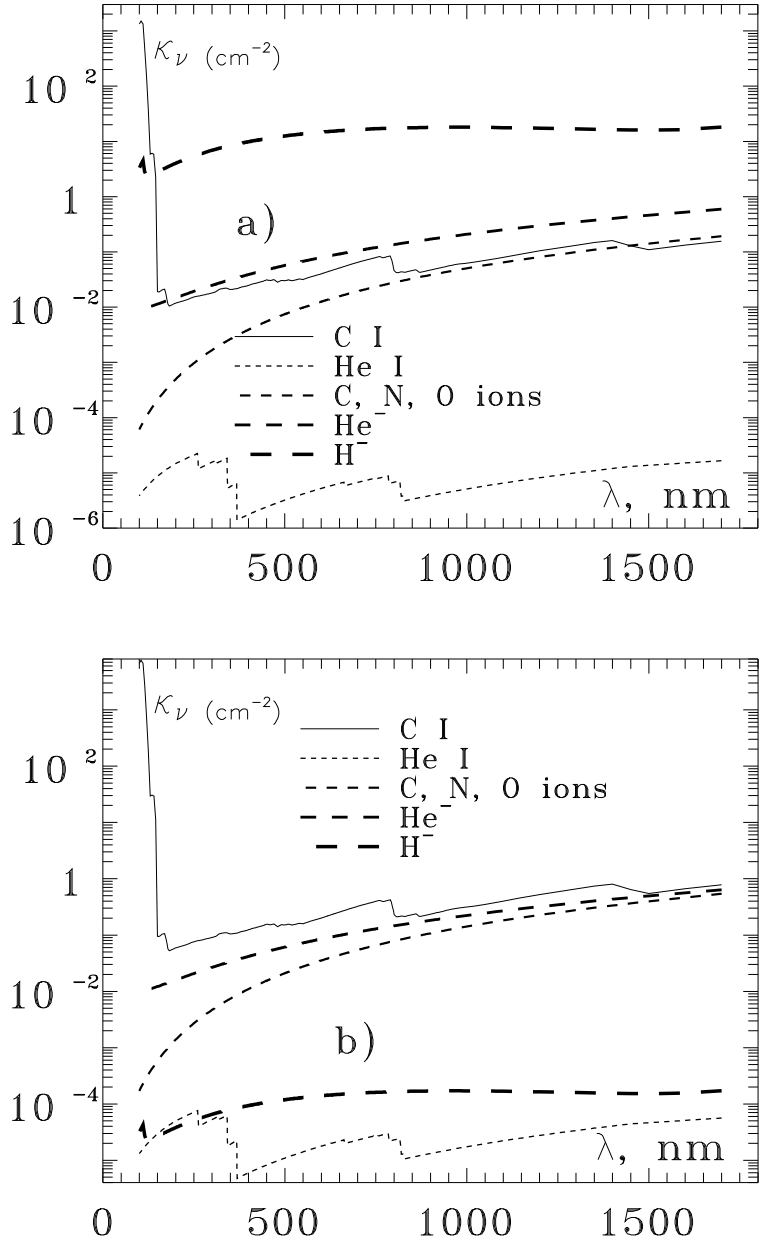}
\caption{Continuum absorption coefficients
computed for solar (top) and R CrB (bottom) model atmospheres.
Computations were carried out for T = 5900 K (Pavlenko 1999).
The most important distinction in Hd atmospheres
is a substantial decrease of H$^-$
absorption and, accordingly, an increase of the importance of
bound-free absorption by neutral carbon atoms proved to
be substantial over large frequency intervals.\label{_OPAC_}}
\end{center}
\end{figure}

At photospheric levels of hotter Hd stars (\Tef $>$ 7000 K), the
continuum opacity  is governed mainly by a bound-free absorption
of C atoms (Fig. \ref{_OPAC_}, see also Pavlenko 1999). To compute
opacities due to bf absorption of C, N, O atoms we used TOPBASE
(Seaton 1982) crossections (Pavlenko \& Zhukovska 2002). At
the same time, opacity above the photosphere of Hd stars with \Tef
$<$ 6500 K is determined, to a large extent, by absorption of 
molecules contained C, N, O atoms (Pavlenko, Yakovina \& Duerbeck 2000).
Naturally, this severely limits the use of ODF-like methods for
the computation of blanketing effects (see discussion in Mihalas
1978). We used JOLA and ``line by line'' models of absorption by
molecules to compute opacities in atmospheres of late-type stars 
for Kurucz (1993), Partrige and
Schwenke (1998), Goorvitch (1994) and Harris (2002) lists.

\section{Dependence $\tau_{ross} = f (T,P)$}
\begin{figure}
\begin{center}
\epsfxsize=100mm \epsfbox{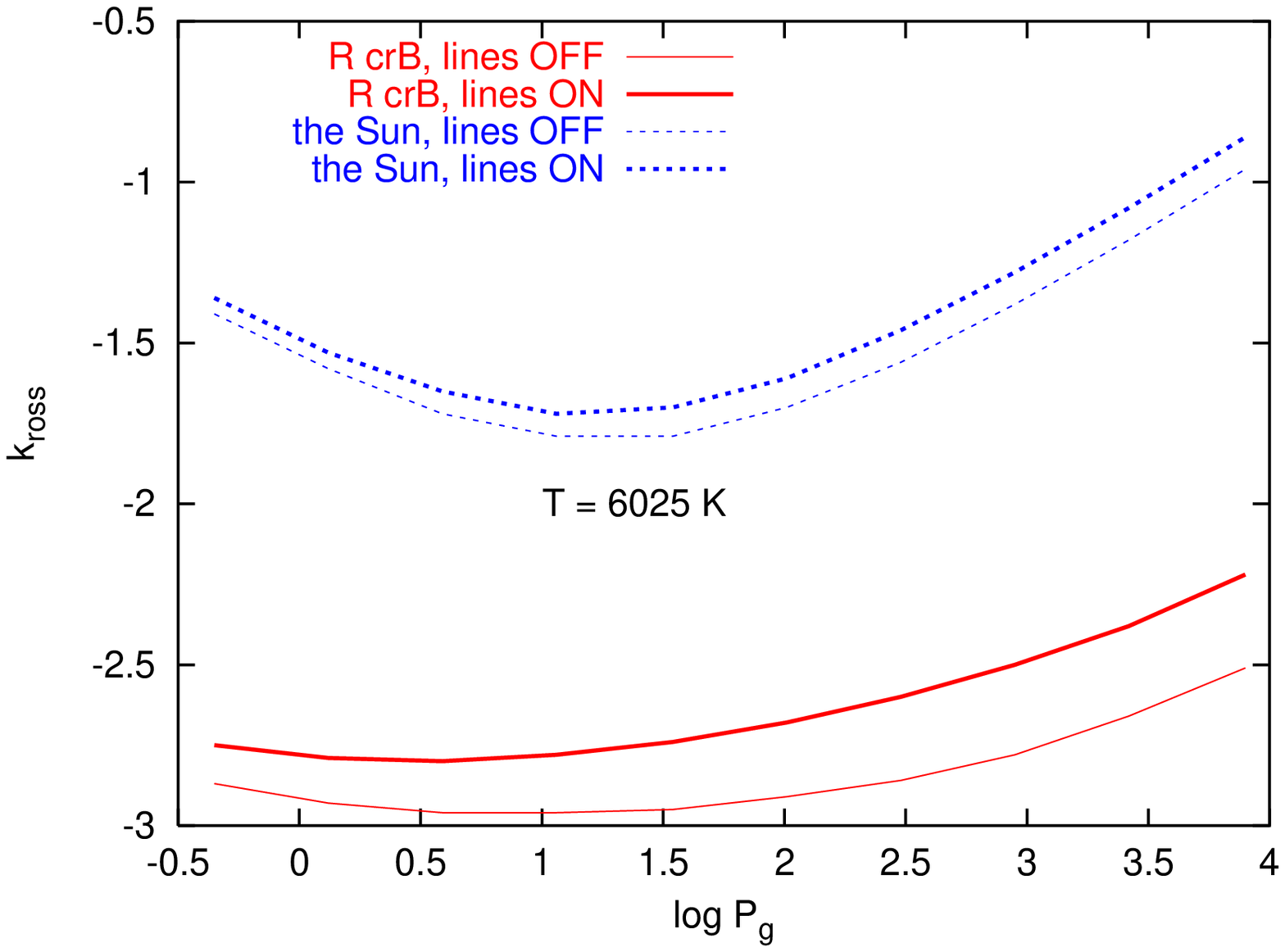}

\epsfxsize=100mm \epsfbox{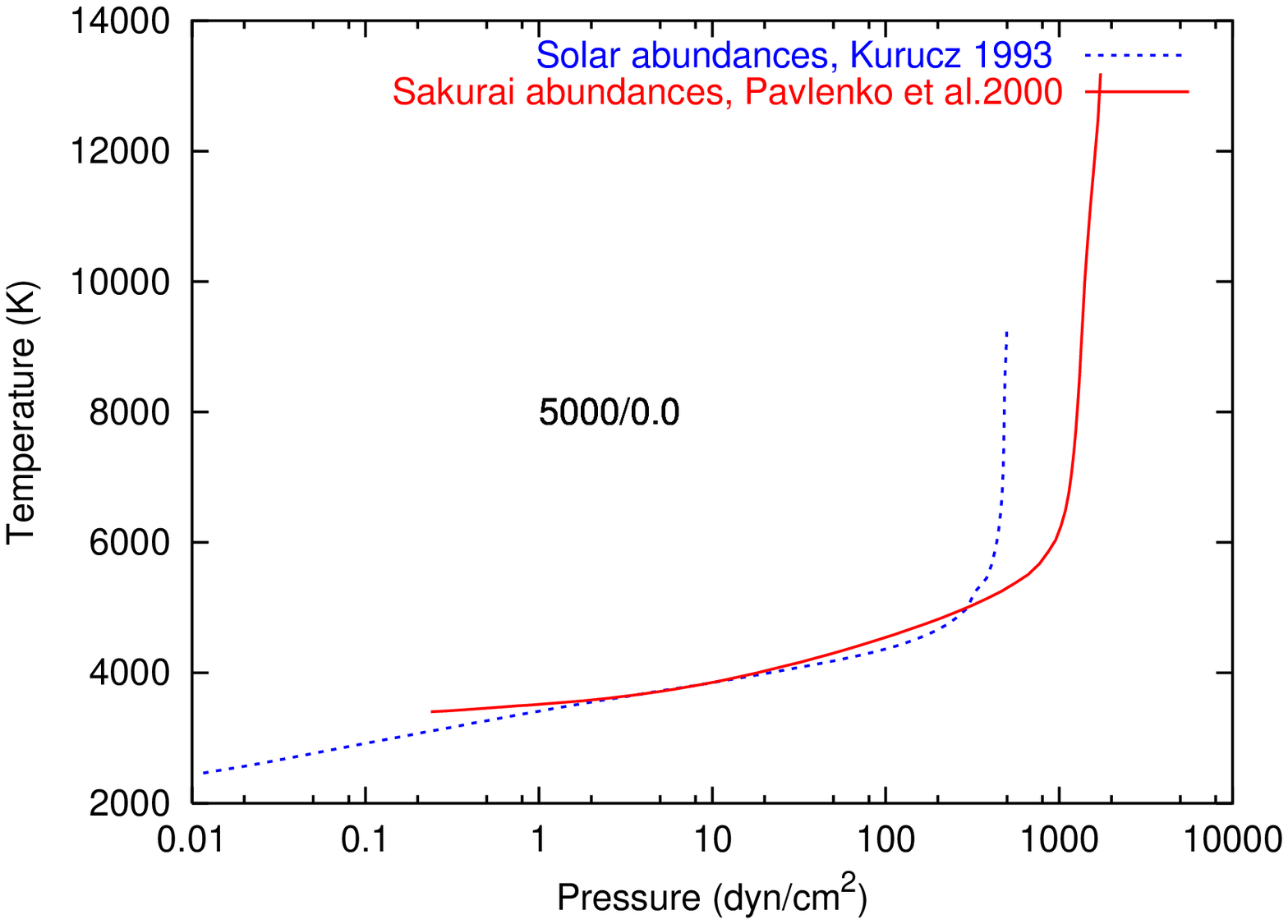}
\caption{Top: absorption coefficients $\kappa_{ross}$ as a
function of gas pressure computed for solar (blue lines) and R CrB
(blue lines) abundances. Computations were carried out for T$_e$
=6000 K. Thin and thick lines indicate the computations with
continuum absorption alone and with continuum + atomic lines,
respectively. Bottom: temperature structures of model atmospheres
5000/0.0 computed for solar and Sakurai's abundances,
respectively.\label{_tauT_}}

\end{center}
\end{figure}

Due to drop of
H$^-$ absorption the mean opacity $\kappa_{ross}$ is reduced in
 photosheres of Hd stars. As result, they are shifted downwards
higher pressure regions (Fig.\ref{_tauT_}).

\section{Model atmospheres}
We computed a grid of stellar model atmospheres  of \Tef =
7000-4000 K, log g = 0 - 1 by SAM12 program (Pavlenko 2002, 2002a).
SAM12 is a modification of ATLAS12 (Kurucz 1999). Opacity
sampling treatment was used to account atomic and molecular
absorption.

\section{Sakurai's object}
\begin{figure}
\begin{center}
\epsfxsize=100mm \epsfbox{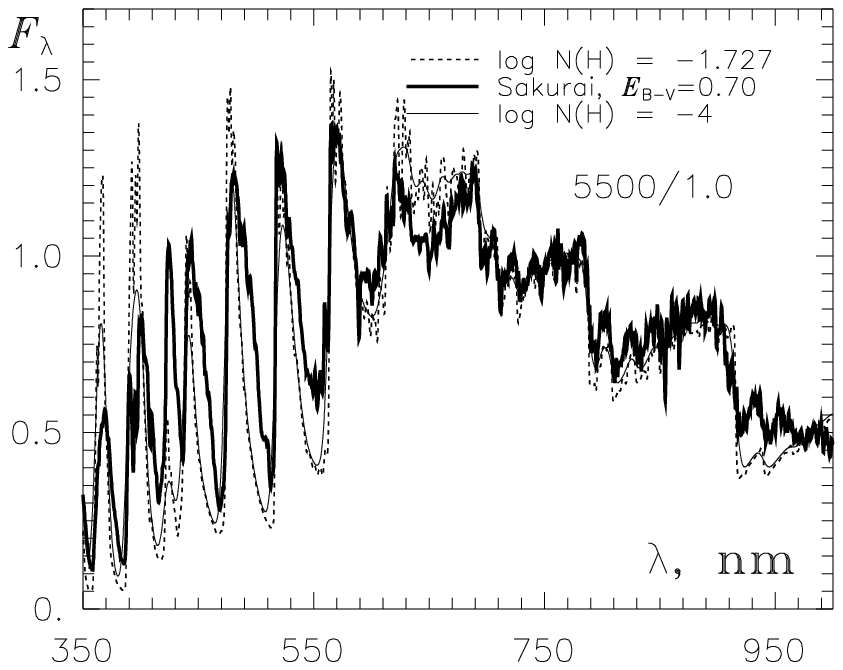}

\epsfxsize=100mm \epsfbox{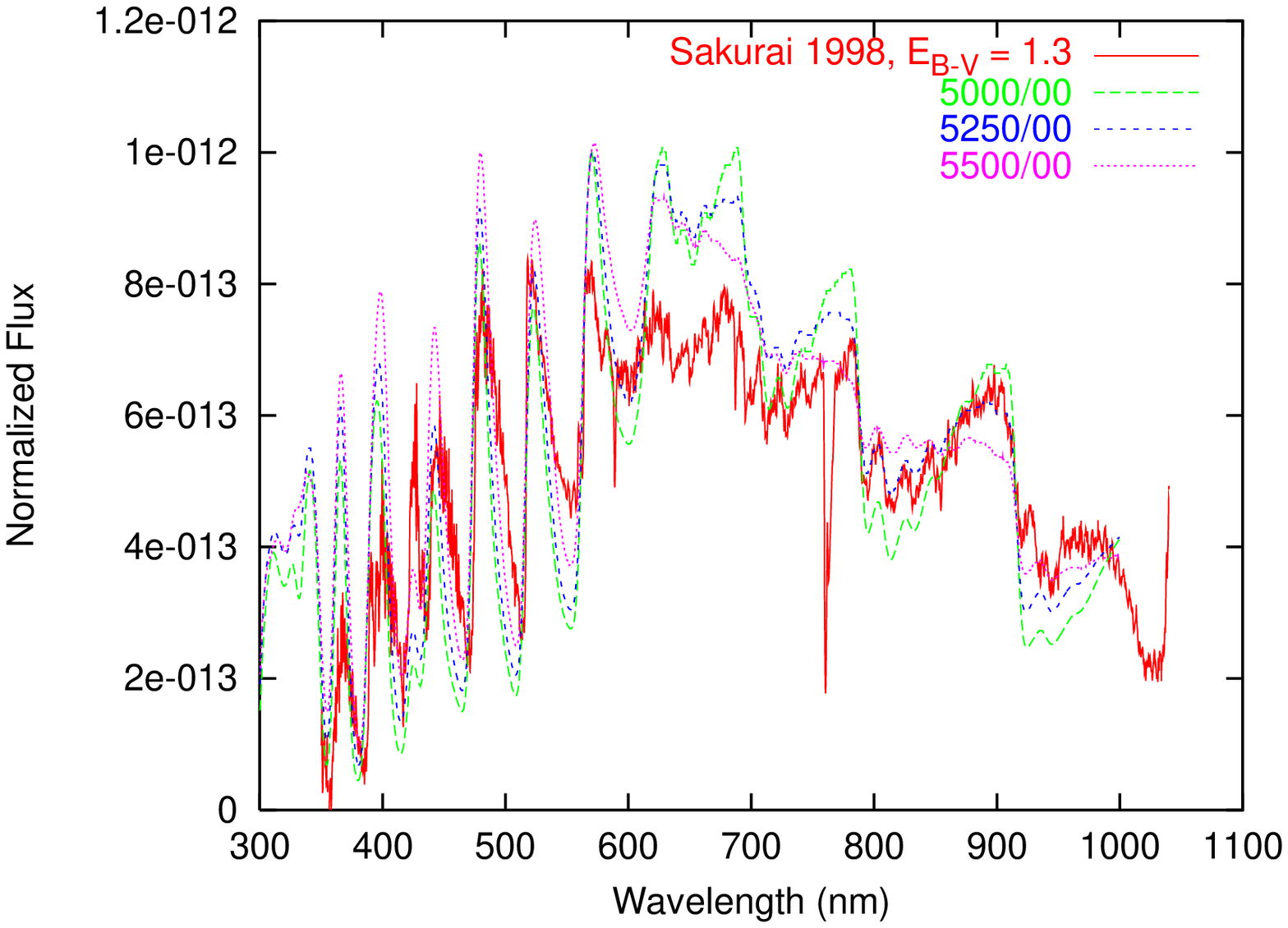}
\caption{Fits to optical SEDs of V4334 Sgr
in 1997-1998.\label{_SOPT_}}
\end{center}
\end{figure}

\begin{figure}
\begin{center}
\epsfxsize=100mm \epsfbox{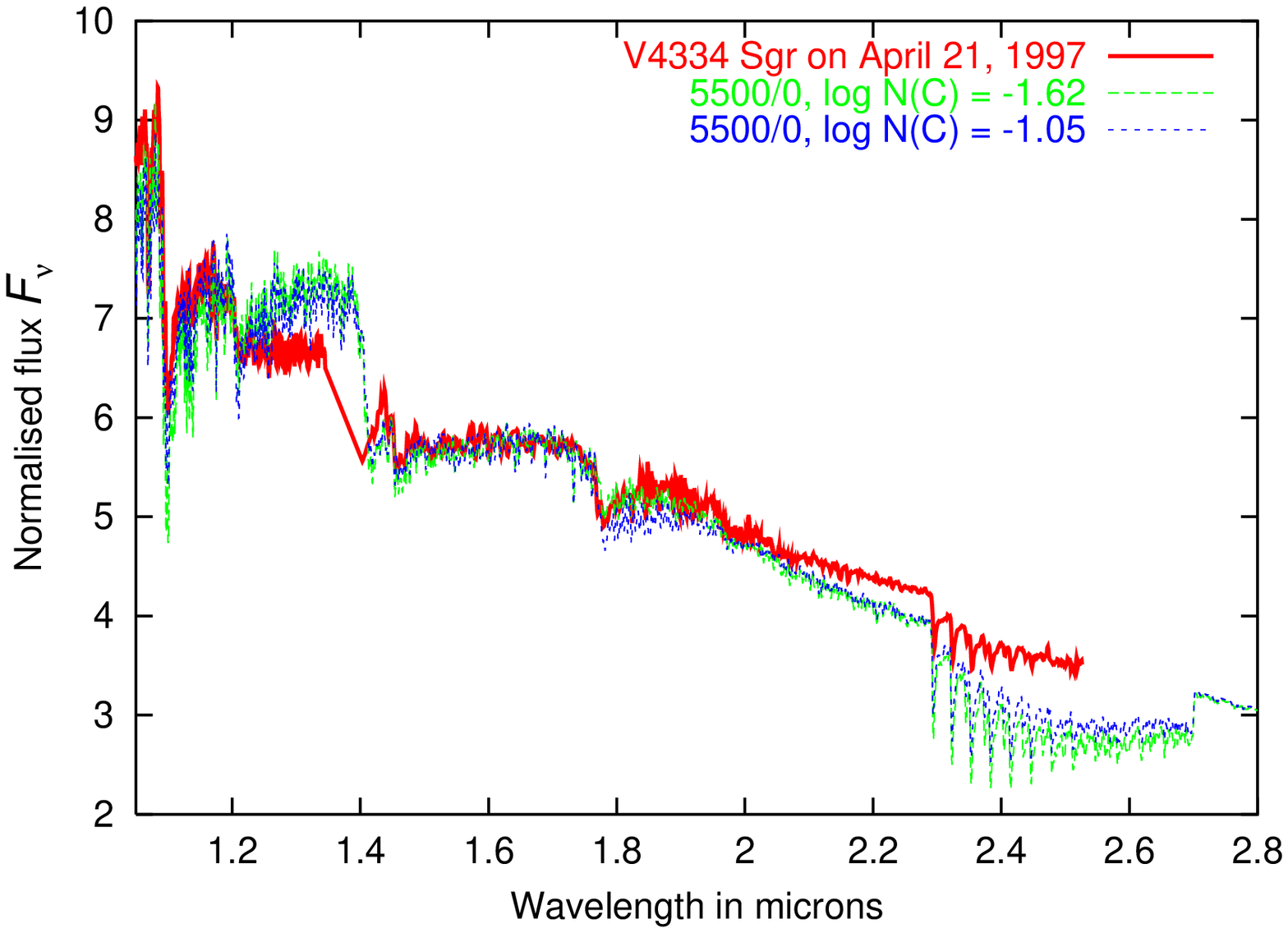}

\epsfxsize=100mm \epsfbox{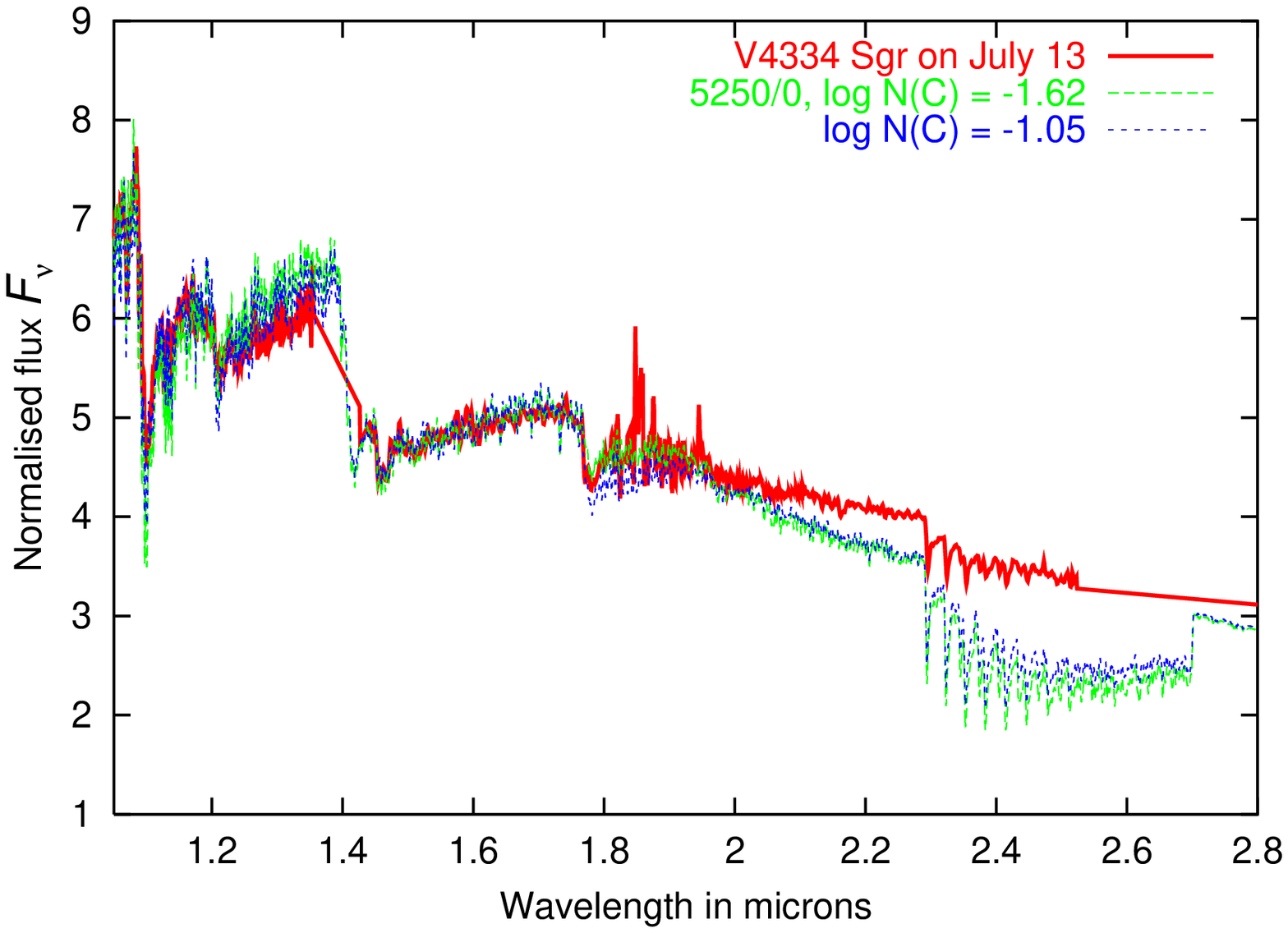}
\caption{Fits to IR SEDs of V4334 Sgr in
1997.\label{_SIR_}}
\end{center}
\end{figure}

Fits of theoretical SEDs to observed in 1997 - 1998 ones allow us 
to determine \Tef and \EBV of Sakurai's object at the latest 
stages of its evolution (Fig. \ref{_SOPT_} in the frame of 
self-consistent approach(Pavlenko et al. 2000, Pavlenko \& 
Duerbeck 2001, Pavlenko \& Geballe 2002).
 Fits to IR spectra allows to clearly determine an 
infrared excess due to emission of hot (T $>$ 1000 K) dust 
(Fig.\ref{_SIR_}, see Pavlenko \& Geballe 2002 for more details).

\section{Przybylski's star}
\begin{figure}
\begin{center}
\epsfxsize=100mm \epsfbox{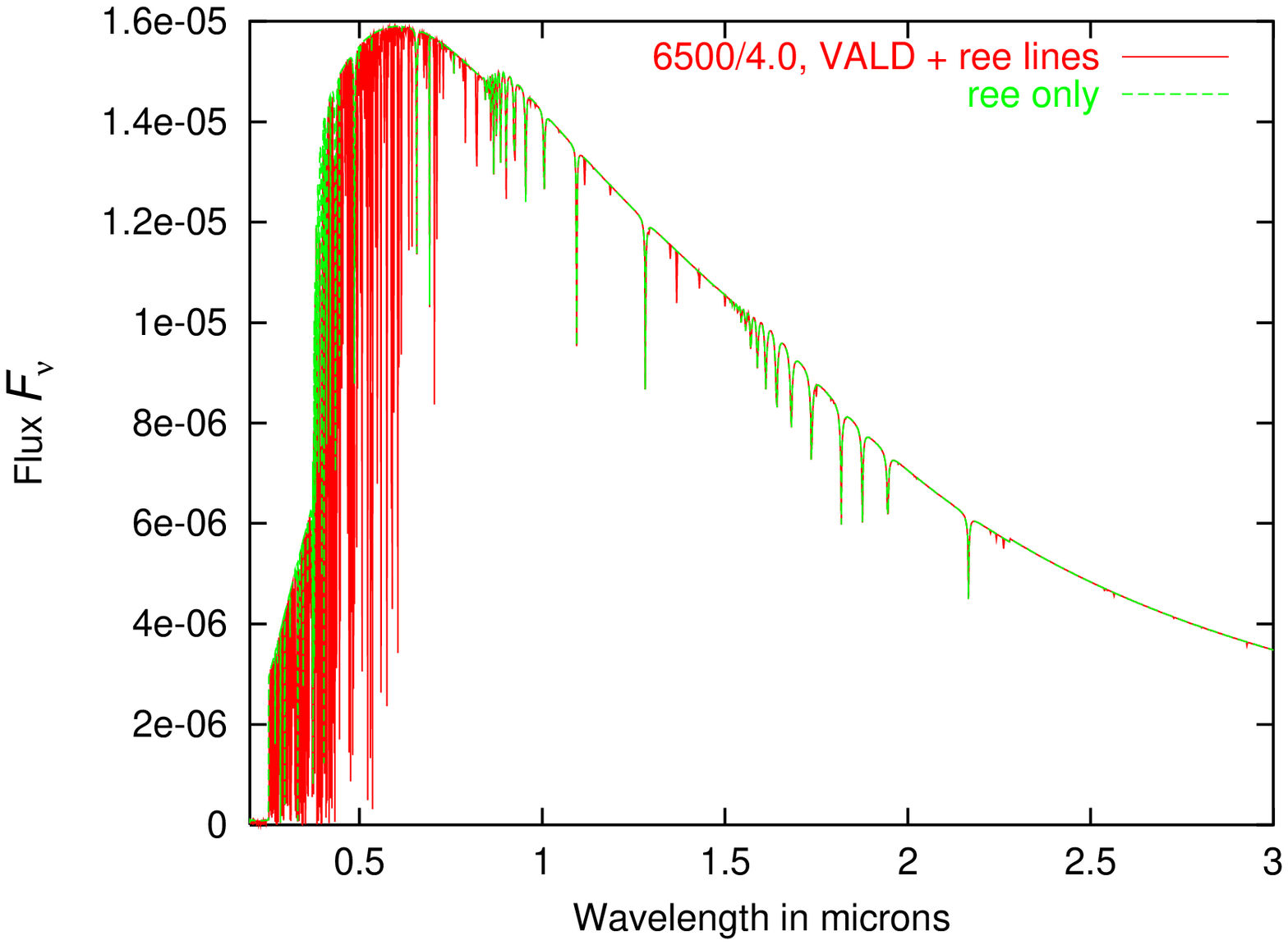}

\epsfxsize=100mm \epsfbox{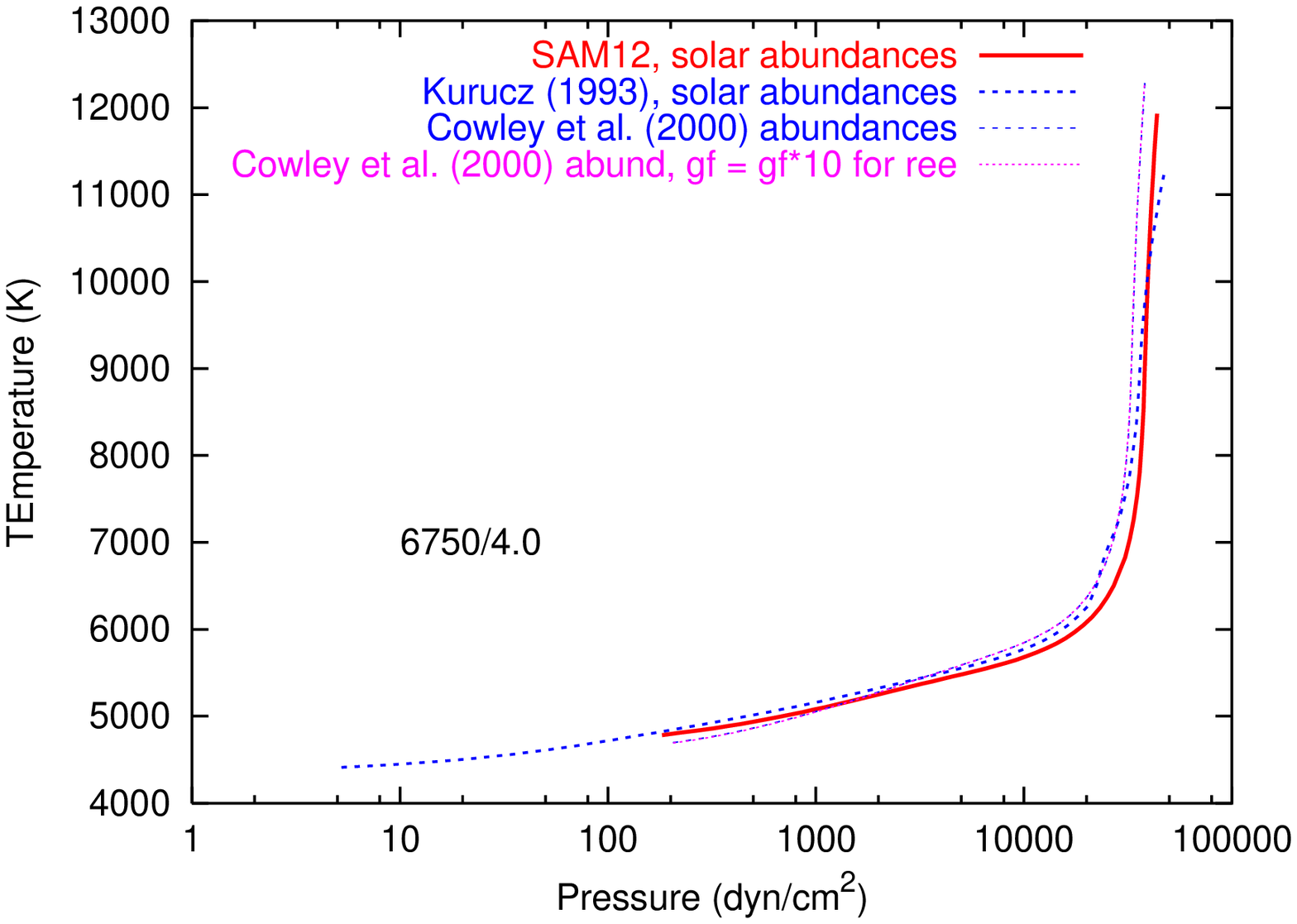}
\caption{Impact of ree absorption on SEDs of
Przybylski's star (top).
Temperature of a few model atmospheres computed for diffrent input
parameters.\label{_PRZ_}}
\end{center}
\end{figure}

Model atmospheres with T$_{eff}$ = 6400-6800~K, $\log g$ =4.0 and
abundances from Cowley et al. (2000) were computed using the
opacity sampling method using VALD (Kupka et al. 1999) and DREAM
(DREAM 2002)
 line lists. Atomic lines are the main source of opacity in
Paczinsky's star (Fig. \ref{_PRZ_}). We found, however, that a
contribution of the bound-bound r-elements into opacity is rather low in
comparison with other elements. The most probably, it is a
consequence of incompleteness of the used line lists of
r-elements.
Computation of complete line lists of r-elements is of crucial
importance for now.

\section{C-giants} 
\begin{figure}
\begin{center}
\epsfxsize=100mm \epsfbox{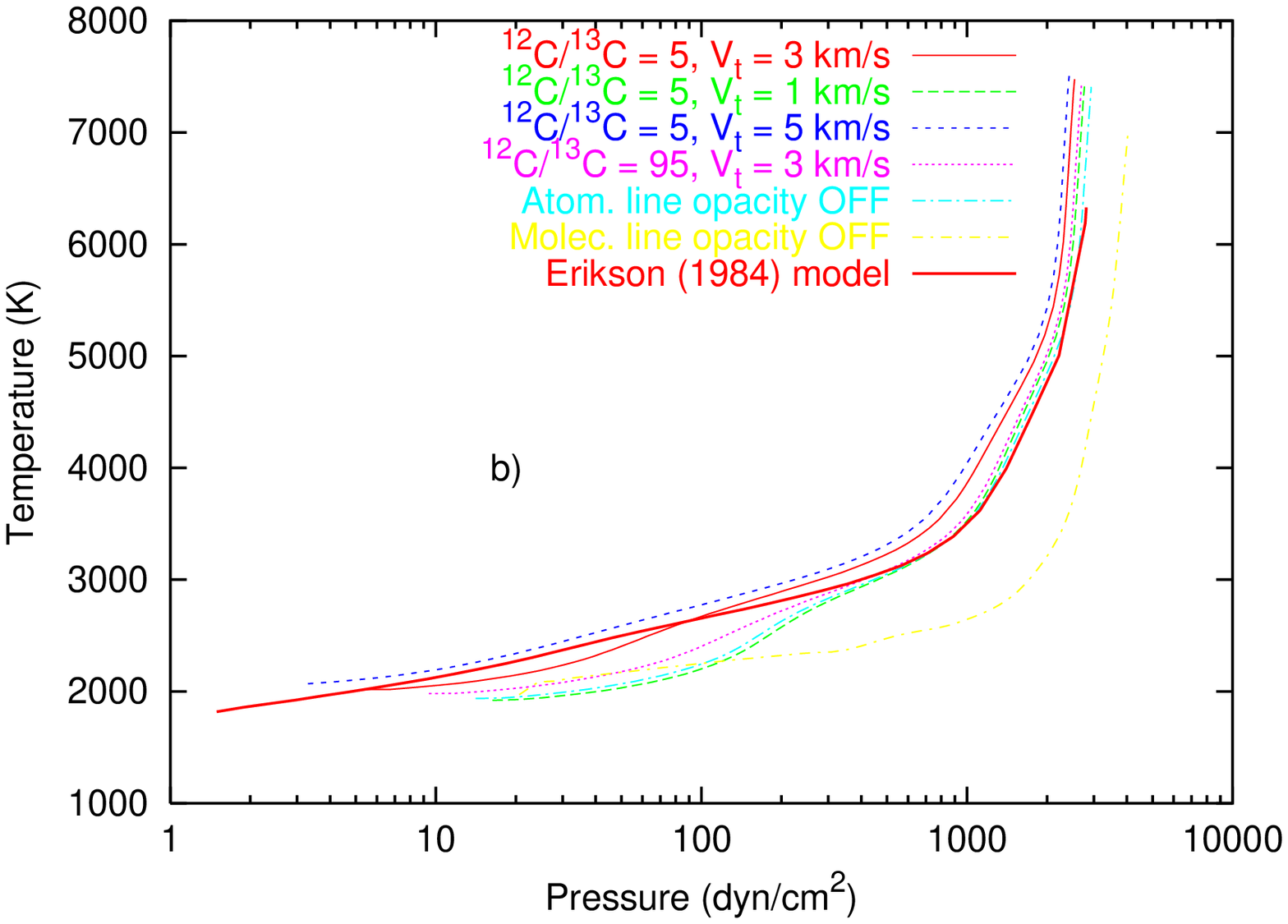}

\epsfxsize=100mm \epsfbox{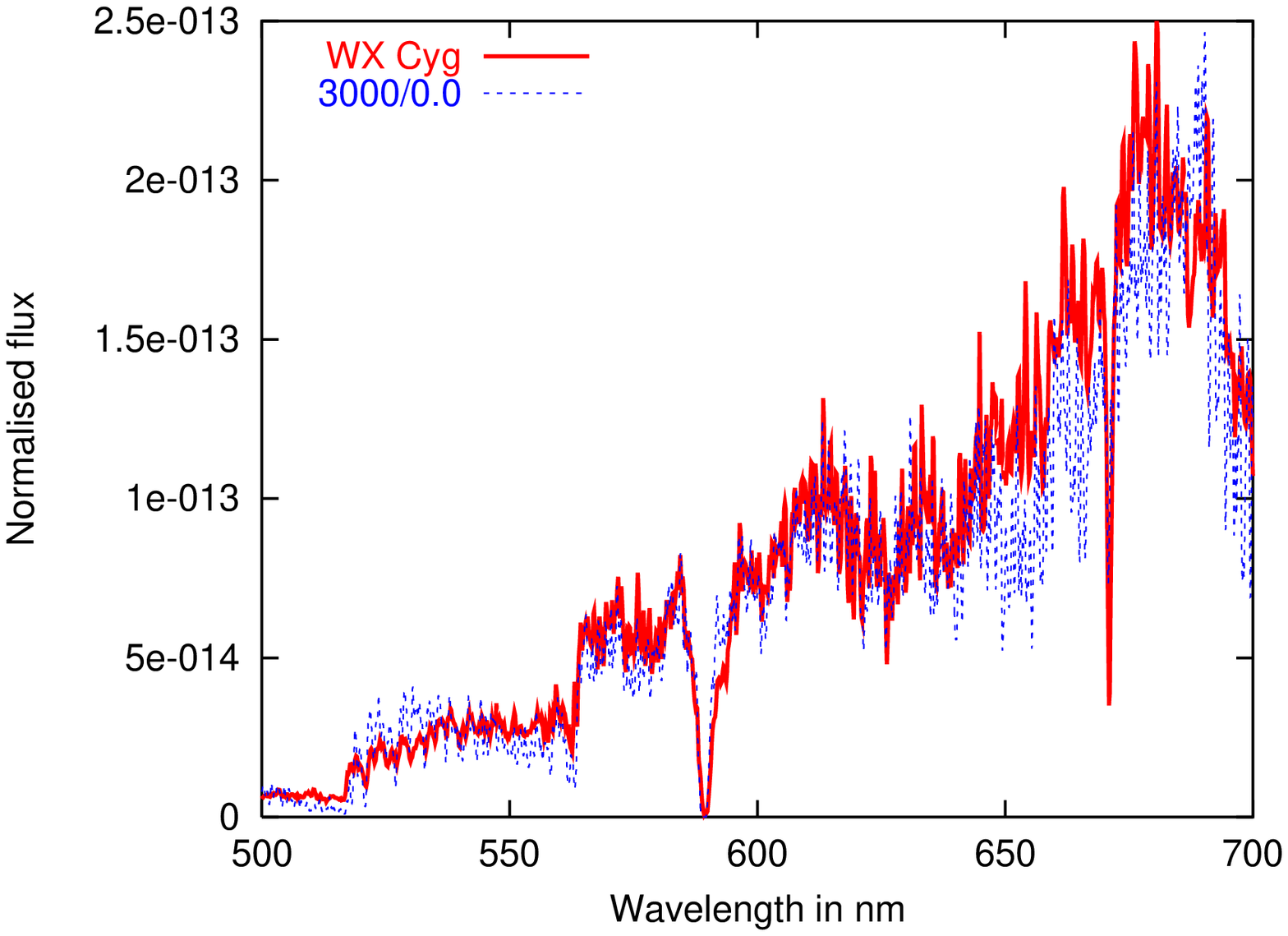}
\caption{Top: dependence of the temperature structure
of
model atmosphere of C-giant
3000/0 on input parameters. Bottom: fit to the observed SED
of WX Cyg (Barnbaum et al. 1996).
\label{_CG_}}
\end{center}
\end{figure}

Two grids of model atmospheres of different
$^{12}$C/$^{13}$C and C/O ratios were computed by SAM12 program
for opacities provided by diatomic molecules (Kurucz 1993) lists
and HCN (Harris 2002). Our model atmospheres of C-giants
(see Pavlenko 2002a) depend on input parameters, therefore fits to
observed spectra are not good enough in some cases (Fig.
\ref{_CG_}).

\acknowledgments

I thank SOC and LOC of the Symposium for financial support of my
participation. I thank Drs. Hilmar Duerbeck, Larisa Yakovina \&
Tomas Geballe for very fruitful collaboration. Partial support of
my investigations was provided by Small Research Grant of AAS.



\newpage

\newpage

\newpage

\newpage

\newpage

\begin{references}
\reference Asplund, M., Gustafsson, B., Kameswara Rao, N., Lambert, D.L.
1998, A\&A, 332, 651.

\reference Barnbaum C., Stone R.P.S., Keenan P.C. 1996.
   ApJ. Suppl. 105, 419.

\reference Cowley, C.R., Ryabchikva, T., Kupka, F.,  Bord, D.J., 
Mathys, G., BIdelman, W.P.. 2000. MNRAS, 317, 299.

\reference Duerbeck. H. 2001. Astroph. Space Sci., in press.

\reference Goorvitch, D., 1994, Astrophys. J. Suppl. Ser., 95, 535.

\reference Harris, G. 2002. PhD Thesis, UCL.

\reference  Kupka, F., Piskunov, N., Ryabchikova, T. A.,
                     Stempels, H. C., Weiss, W. W. 1999.
                  A\&A Suppl., 138, 119.


\reference Kurucz R.L., 1993, CD-ROMs, Cambridge, Harvard Univ.


\reference Kurucz, R.L. 1999. http://cfa5.harvard.edu

\reference Mihalas. D. 1978. Stellar atmospheres. Freeman and Co. San
Francisco

\reference Partridge H., Schwenke D.W., 1997, J.Chem. Phys., 106, 4618

\reference Pavlenko, Ya., Yakovina, L. 1994, Astr. Reports, 38, 768

\reference Pavlenko, Ya. 1999, Astr. Reports, 43, 94

\reference Pavlenko, Ya.,Yakovina, L., Duerbeck, H.W. 2000, A\&A, 354, 229



\reference Pavlenko, Ya.,Duerbeck, H.W. 2001, A\&A, 367, 933.

\reference Pavlenko, Ya. 2002. Astron. Repts, {\em submitted}

\reference Pavlenko, Ya. 2002a. http://www.mao.kiev.ua/staff/yp/TOP-mod.htm

\reference Pavlenko, Ya. 2002b. Ap\&SS, 279, 91 

\reference Pavlenko, Ya., Geballe T.,  2002. A\&A, 390, 621.  

\reference Pavlenko, Ya., Zhukovska, S. KFNT, {\em submitted}

\reference Seaton, M.J. 1992
Rev. Mexicana Astron. Astrofis. 23, 180.

\end{references}
\end{document}